\documentclass[twocolumn,english,aps,pra]{revtex4-1}
\usepackage[T1]{fontenc}
\usepackage[latin9]{luainputenc}
\usepackage{amsmath}
\usepackage{amssymb}
\usepackage{graphicx}
\usepackage{float}
\usepackage{xcolor}



\usepackage{babel}
\begin{document}
\title{Coherence behavior of strongly coupled bosonic modes}
\author{Jucelino F. Sousa$^{1,4}$, Carlos H. S. Vieira$^{2}$, Jonas F. G. Santos$^{3}$, and Irismar G. da Paz$^{4}$}
\affiliation{$^{1}$Departamento de F\'{i}sica, Universidade Federal do Maranh\~{a}o, Campus Universit\'{a}rio do Bacanga, 65080-805, S\~{a}o Lu\'{i}s, Maranh\~{a}o, Brazil.}
\affiliation{$^{2}$Centro de Ci\^{e}ncias Naturais e Humanas, Universidade Federal do ABC, Avenida dos Estados 5001, 09210-580, Santo Andr\'{e}, S\~{a}o Paulo, Brazil}
\affiliation{$^{3}$Division of Natural Sciences, Duke Kunshan University, No. 8 Duke Avenue, Kunshan, Jiangsu 215316, China.}
\affiliation{$^{4}$ Departamento de F\'{i}sica, Universidade Federal do Piau\'{i}, Campus Ministro Petr\^{o}nio Portela, CEP 64049-550, Teresina, PI, Brazil}

\begin{abstract}
We study the effect of the intermode coupling in the generation of coherence when two bosonic modes are bilinearly coupled. We consider the case for which the two modes are weakly coupled and the rotating-wave approximation (RWA) applies and the case for which they are strongly coupled and the (RWA) does not apply. Then, in the regime of validity of (RWA), there is no coherence generation solely due to squeezing effects, which means there is an exchange of excitation between the modes and negligible squeezing. On the other hand, if the two modes are strongly coupled coherence is generated by the squeezing interaction. For the system of two bosonic modes weakly coupled with a Markovian bath at temperature $T$ the coherence decreases with $T$. In general, when both kinds of couplings (exchange of excitation and squeezing) are turned on the exchange of excitation contributes to generating more coherence in comparison with a purely squeezing coupling. Thus, the coherence decreases more slowly with the temperature when both couplings are present. Finally, we explore the case when only one of the two coupled modes interacts with a Markovian bath while the other one remains free from the environment. We observe that the intermode coupling induces oscillations in the coherence and fidelity dynamics similar to the behavior for the coupling with a non-Markovian environment. 
\end{abstract}

\maketitle

\section{Introduction}
The recent developments of the so-called quantum science and technology have pave the way for current and future applications of quantum resources with no classical counterpart as entanglement and coherence. There are important different applications, for example, in quantum metrology \cite{Scorpo2017, Yuan2020, Ge2020}, quantum computation \cite{Arute2019, Gedik2015} and many other \cite{Dorfman2013, Klatzow2019, Aasi2013, Bermudez2013, Lin2021, Sheng2021, Dieguez2022}. Concurrently, the ability to experimentally access quantum signatures has expanded considerably, allowing for test-of-principle verification for single \cite{Peterson2019, Batalhao2014, Micadei2019} or collective \cite{Riofrio2017, Klatzow2019,Maslennikov2019} quantum systems. Examples of experimental platforms for this purpose are quantum optical devices and trapped ions systems, where it is possible to work with continuous or discrete variables \cite{Cai2021, Ren2021}. In both of them, quantum coherence has played a fundamental role. With no classical counterpart, quantum coherence has being recently investigated in the scope of thermodynamics, related with the entropy production in quantum processes \cite{Santos2019, Santos2021A} or even in cyclic operations \cite{Camati2019, Dann2020}. Furthermore, coherence has demonstrated an important aspect in fundamental issues, as in quantum phase transition \cite{Rossatto2020, Li2016}.

There are different methods to generate coherence in the state of quantum systems. A general way is related with the finite-time dynamics in a unitary process. In particular, for harmonic oscillators systems, a finite-time unitary process changing the frequency is always associated with the production of coherence in the energy basis \cite{Graham1987}, which the finite-time properties are captured by the well-known Hussimi parameter $Q^\star$ \cite{Husimi1953, Rossnagel2014, Abah2020}. Another relevant manner to produce coherence is employing squeezing operations \cite{Graham1987, Galve2009}, with this technique experimentally robust \cite{Xin2021}. In fact, squeezing has been employed to protect the coherence of quantum system \cite{Marinho2020}. In this sense, given a quantum system subject to a general dynamics, it is important to understand the relation between the degree of squeezing and the amount of coherence of its state.

It was shown that modeling the master equation of two coupled modes interacting with a Markovian bath by using local Lindblad terms is valid only if the intermode coupling is weak \cite{Joshi2014}. When the intermode coupling is strong the master equation contains nonlocal Lindblad terms. It was also shown that the expression to the steady state is different when the intermode are strongly coupled. The behavior of the entanglement for the steady-state was shown to be changed by the non-local Lindblad terms. In addition, the quantum fidelity between the two one-mode states has been analyzed in order to compare the difference in the dynamics of the local and non-local Lindblad approaches \cite{Joshi2014}.

This work is focused on the role of the intermode coupling in the generation and dynamics of coherence in the energy basis for two-coupled bosonic modes. This paradigmatic setup represents many suitable scenarios in quantum physics, for instance, the hopping coupling between quantum resonators, with applications in quantum phase transitions and quantum computation \cite{Wang2016, Wang2020}, the use as working substance in quantum heat machines model \cite{Sheng2021}, and the Dicke model for large number of spins \cite{Emary2003}. With this purpose, we consider three possible situations. The first one corresponds to two coupled bosonic modes free of any environment interaction and subject to a unitary evolution. In the second case we assume that both bosonic modes interact with identical environments. More precisely, it is considered a Markovian thermal bath. In the last configuration only one mode interacts with the bath, whereas the other plays the role of an auxiliary system (ancilla), and we show that depending on the choice of parameters, the evolution of the system can be understood as Markovian or non-Markovian.  

Our treatment of the Hamiltonian of two-coupled modes interacting with a general coupling in principle, allow to consider three physical regimes, i.e., one in which there is only exchange of excitation, one in which there is only squeezing and one in which both effects are present. By considering the relation between the parameters,  the first case can be a weak intermode squeezing coupling and the (RWA) applies, which also means negligible squeezing. In this case we do not have coherence generation due exclusively to squeezing effects. The last two cases represent a strong intermode coupling and the (RWA) is no longer valid. Proceeding in this way is useful to properly understand the effect of the intermode coupling in the coherence behavior. We stress that in the three regimes the initial state is set to be the ground state.

The organization of the work is the following. In Sec. \ref{section01} we introduce the general model and also the measure to quantify the coherence. It must be noted that our system will be always Gaussian, which motivates the use of a measure based on the covariance matrix. The Sec. \ref{section02} is dedicated to explore the role of the intermode coupling in the coherence generation when the system interacts with Markovian thermal baths. In Sec. \ref{section03} we study the effect of the intermode coupling in the local coherence dynamics when only one of the modes interacts with a Markovian thermal bath. Conclusions and final discussions are draw in Sec. \ref{Conclusion}.

\section{Coherence of two-coupled bosonic modes} \label{section01}

In this section, we introduce the model composed of two bosonic modes, each one with natural frequency $\omega$ and coupled by mean of a general bilinear interaction. The Hamiltonian of the full system reads
\begin{equation}
\hat{\mathcal{H}}_S=\omega\hat{a}^{\dagger}\hat{a}+\omega\hat{b}^{\dagger}\hat{b} + U\left(\hat{a},\hat{a}^\dagger,\hat{b}, \hat{b}^\dagger\right), \label{eq:5}
\end{equation}
with $\hat{a}$ ($\hat{a}^\dagger$) and $\hat{b}$ ($\hat{b}^\dagger$) the annihilation (creation) operators for each mode and $U = U\left(\hat{a},\hat{a}^\dagger,\hat{b}, \hat{b}^\dagger\right)$ represents a general coupling between them. 

This particular model for the coupling is relevant, for instance, to describe the well-known Dicke model when the number of spins is very large \cite{Emary2003}. Also, it is used to describe a one-dimension bosonic modes chain, which has fundamental relevance in thermal transport \cite{Nicacio2014}. Furthermore, depending on the choice for the coupling $U$ the interaction can describe change of excitations between the modes or two-mode squeezing. In order to capture all the relevant effects we are interested, we set the interaction to be
\begin{equation}
U = \lambda\left(\hat{a}^\dagger \hat{b} + \hat{a}\hat{b}^\dagger\right) + \mu\left(\hat{a}^{\dagger}\hat{b}^{\dagger}+\hat{a}\hat{b}\right).
\label{coupling}
\end{equation}
Here, the parameters $\mu$ and $\lambda$ tune the strength of the two-mode squeezing and the exchange of excitation between the two bosonic modes, respectively.  For $\mu\ll\omega$ the squeezing coupling is weak, one can neglect its contribution, i.e., $\mu=0$ and the (RWA) applies. In fact, (RWA) approximation means dropping quickly oscillating terms from the Hamiltonian. Therefore, for $\mu\ll\omega$ the two-mode squeezing terms $\hat{a}\hat{b}$ and $\hat{a}^{\dagger}\hat{b}^{\dagger}$ can be neglected since they oscillate, in the interaction picture, on a scale $\sim \omega^{-1}$ which is much faster than the time scale $\mu^{-1}$ \cite{Schleich}. On the other hand, for $\mu\sim \omega$ the squeezing coupling is strong, one can not neglect its contribution, i.e., $\mu\neq0$ and the (RWA) does not apply.

 The system is illustrated in Fig. 1, where $\mathcal{F}=\mathcal{F}(\lambda,\mu)$ is the intermode coupling. This paradigmatic model allows to study one-mode squeezing (local effects) associated with coherence in the local energy basis, as well as two-mode squeezing (global effects) which is connected with entanglement and global coherence.  We are interested in studying the effect of strong intermode coupling on the behavior of the coherence. Then, we will consider the case of purely squeezing intermode coupling (when $\lambda=0$), purely exchange of excitation intermode coupling (when $\mu=0$), as well as the case where both effects are present. 
\begin{figure}[H]
\centering
\includegraphics[scale=0.4]{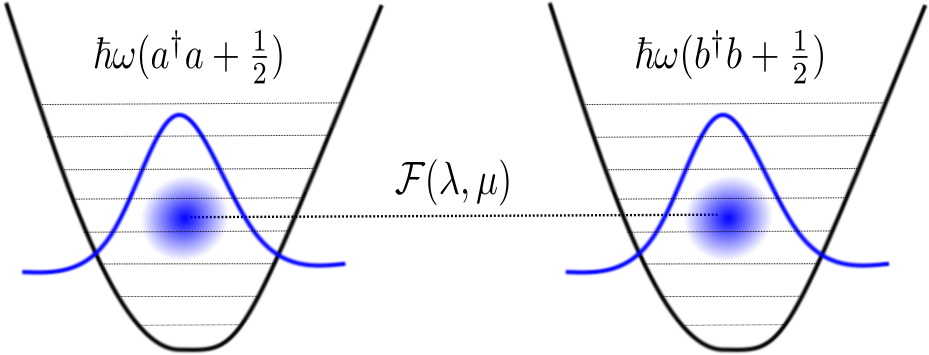}
\caption{The basic setup. Two bosonic modes interacting through coupling Hamiltonian in Eq. (\ref{coupling}). We highlight the importance of this kind of system in the study of the Dicke model and in thermal transport.}
\end{figure}

In order to study the coherence dynamics of the model, we assume the system is initially in the ground state. The main motivation is that it is sufficient to derive all results, as well as we avoid considering other relevant effects such as non-Gaussian properties. Also, ground states are easier to experimentally prepare than excited states.
The wave function of the ground state in terms of the position variable $x_{a,b}$ is given by (see Appendix A)
\begin{align}
\Psi(x_{a},x_{b}) & =\frac{2\sqrt[4]{\kappa_{1}\kappa_{2}}}{\sqrt{\pi}}\cdot\exp{\left[-\kappa_{1}(x_{a}+x_{b})^{2}\right]}\nonumber \\
 & \times \exp{\left[-\kappa_{2}(x_{a}-x_{b})^{2}\right]},\label{eq:16}
\end{align}
with 
\begin{equation}
\kappa_{1}=\frac{\omega\sqrt{\omega+\lambda+\mu}}{4\cdot\sqrt{\omega+\lambda-\mu}}\; ,\kappa_{2}=\frac{\omega\sqrt{\omega-\lambda-\mu}}{4\cdot\sqrt{\omega-\lambda+\mu}}.\label{eq:17}
\end{equation}
One can observe that if $\mu=0$, $\kappa_{1}=\kappa_{2}=\omega/4$, i.e., they are $\lambda$ independent and the off-diagonal elements of the covariance matrix in Eqs. (\ref{eq:20}) and (\ref{eq:22}) are identically zero. Furthermore, in order to consider stable solutions, we impose the condition $2\lambda < \omega$ (see Appendix A).

The quantification of coherence in operational terms has been established in Ref. \cite{Baumgratz2014}, and it is expressed using the relative entropy. It is valid for any kind of state, in continuous or discrete variables. In particular, for continuous variables systems under a class of dynamics such that the state is Gaussian for any time, we can use an extension of Ref. \cite{Baumgratz2014} which uses the covariance matrix to quantify the coherence \cite{Xu2016}. This is the case in the three possible configurations we assume below. For this purpose, one consider the relative entropy for two distinct states, $\rho$ and $\zeta$, $\mathcal{C}(\rho||\zeta) = \mathrm{Tr}\left[\rho\ln\rho\right]-\mathrm{Tr}\left[\rho\ln\zeta\right]$. Then, we minimize $\mathcal{C}(\rho||\zeta)$ over all the set of incoherent Gaussian states $\zeta$, which are thermal states \cite{ Xu2016}. After this mathematical calculation the measure of coherence for a N-mode Gaussian state is \cite{Baumgratz2014, Xu2016}
\begin{equation}
\mathcal{C}[\rho(\sigma,\vec{d})]=\mathcal{S}(\zeta)-\mathcal{S}(\rho),\label{eq:1}
\end{equation}
where $\sigma$ and $\vec{d}$ are the associated covariance matrix and displacement vector, respectively, $\zeta$ is now the reference thermal state, and we rewritten  $\mathcal{C}[\rho(\sigma,\vec{d})]$ as 
\begin{equation}
\mathcal{C}[\rho(\sigma,\vec{d})]=-S(\rho)+\sum_{j=1}^{N}\left[(\bar{\varepsilon}_{j}+1)\ln\left(\bar{\varepsilon}_{j}+1\right)-(\bar{\varepsilon}_{j})\ln\left(\bar{\varepsilon}_{j}\right)\right],\label{eq:3}
\end{equation}
with the average number of excitations of the reference state is
\begin{equation}
\bar{\varepsilon}_{j}=\frac{1}{4}\left[\sigma_{11}^{j}+\sigma_{22}^{j}+\left(d_{1}^{j}\right)^{2}+\left(d_{2}^{j}\right)^{2}-2\right],\label{eq:4}
\end{equation}
and the von Neumann entropy $S(\rho)$ reads
\begin{equation}
\mathcal{S}(\rho)=-\sum_{j=1}^{N}\left[\frac{\nu_{j}-1}{2}\ln\left(\frac{\nu_{j}-1}{2}\right)-\frac{\nu_{j}+1}{2}\ln\left(\frac{\nu_{j}+1}{2}\right)\right],\label{eq:2}
\end{equation}

where $\lbrace{\nu_j \rbrace}_{j=1}^N$ are the symplectic eigenvalues of $\sigma$ \cite{new00}. The relative entropy of coherence also has been used to quantify quantum correlations in bipartite systems \cite{new06}.

We now move to investigate the ground-state coherence generated by the Hamiltonian in Eq.(\ref{eq:5}), and how the coupling parameters $\lambda$ and $\mu$ affect the coherence generation. The covariance matrix and its elements for the ground state are given by

\begin{equation}
\boldsymbol{\sigma}=\left(\begin{array}{cccc}
\left\langle \hat{x}_{a}^{2}\right\rangle  & 0 & \left\langle \hat{x}_{a}\hat{x}_{b}\right\rangle  & 0\\
0 & \left\langle \hat{p}_{a}^{2}\right\rangle  & 0 & \left\langle \hat{p}_{a}\hat{p}_{b}\right\rangle \\
\left\langle \hat{x}_{a}\hat{x}_{b}\right\rangle  & 0 & \left\langle \hat{x}_{b}^{2}\right\rangle  & 0\\
0 & \left\langle \hat{p}_{a}\hat{p}_{b}\right\rangle  & 0 & \left\langle \hat{p}_{b}^{2}\right\rangle 
\end{array}\right),\label{eq:18}
\end{equation}
\begin{equation}
\left\langle \hat{x}_{a}^{2}\right\rangle =\left\langle \hat{x}_{b}^{2}\right\rangle =\frac{1}{16}\left(\frac{1}{\kappa_{1}}+\frac{1}{\kappa_{2}}\right),\label{eq:19}
\end{equation}
\begin{equation}
\left\langle \hat{x}_{a}\hat{x}_{b}\right\rangle =\frac{1}{16}\left(\frac{1}{\kappa_{1}}-\frac{1}{\kappa_{2}}\right),\label{eq:20}
\end{equation}
\begin{equation}
\left\langle \hat{p}_{a}^{2}\right\rangle =\left\langle \hat{p}_{b}^{2}\right\rangle =\left(\kappa_{1}+\kappa_{2}\right),\label{eq:21}
\end{equation}
and
\begin{equation}
\left\langle \hat{p}_{a}\hat{p}_{b}\right\rangle =\left(\kappa_{1}-\kappa_{2}\right).\label{eq:22}
\end{equation}

Next, we can use the Eq.(\ref{eq:3}) to compute the two-mode quantum coherence (see details in Appendix A). In Fig. \ref{freecase} we show the quantum coherence as a function of the intermode coupling. In Fig. \ref{freecase}-(a) we show the coherence as a function of the squeezing parameter $\mu$ for three values of the exchange of excitation parameter $\lambda$. In Fig. \ref{freecase}-(b) we show the coherence as a function of the exchange of excitation parameter for three values of the squeezing parameter. We can observe that there is coherence production for strong squeezing intermode coupling. On the other hand, for null squeezing, there is no coherence production solely to squeezing effects even if the intermode coupling (exchange of excitation) is strong (when $\lambda \sim \omega$). 
\begin{figure}[H]
\centering
\includegraphics[scale=0.3]{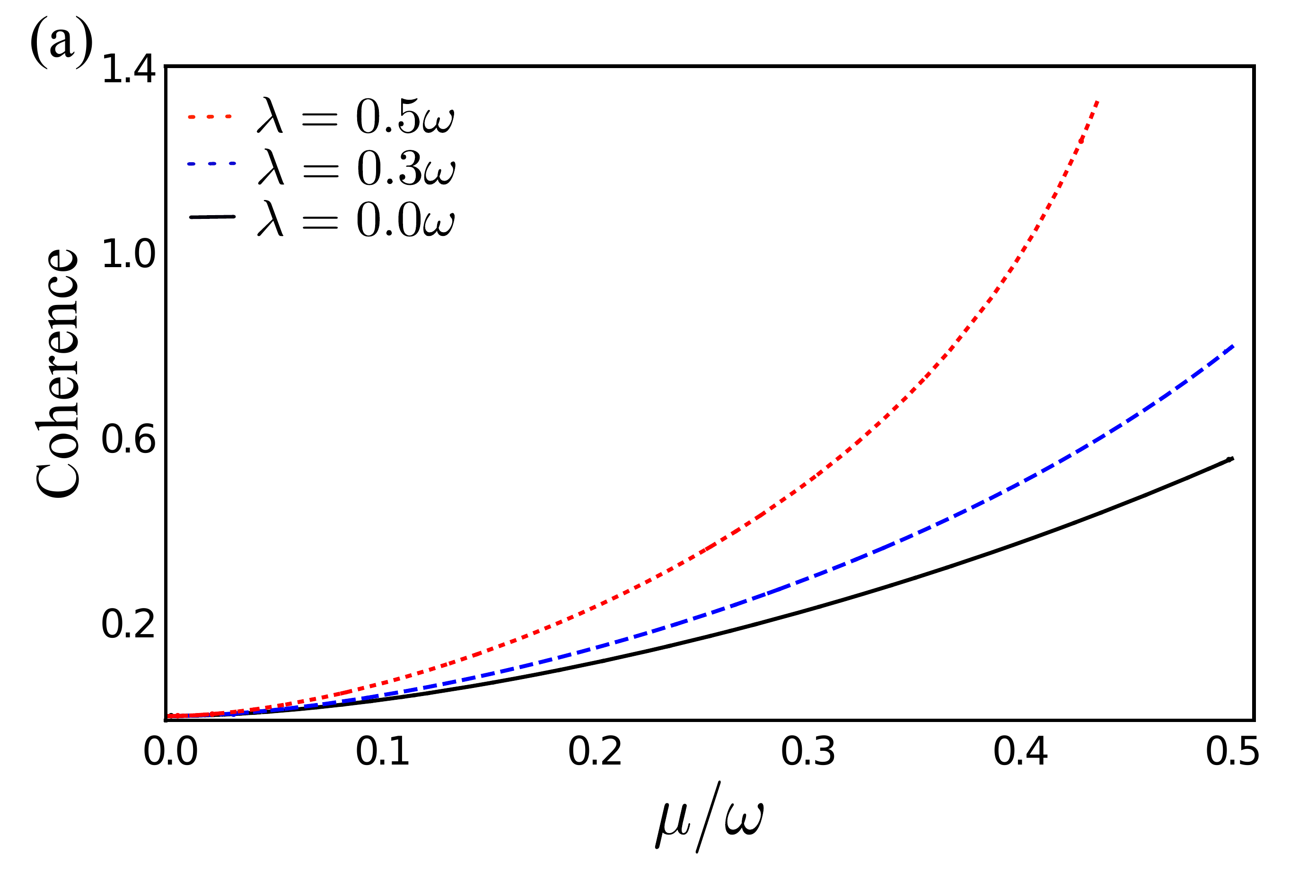}
\includegraphics[scale=0.3]{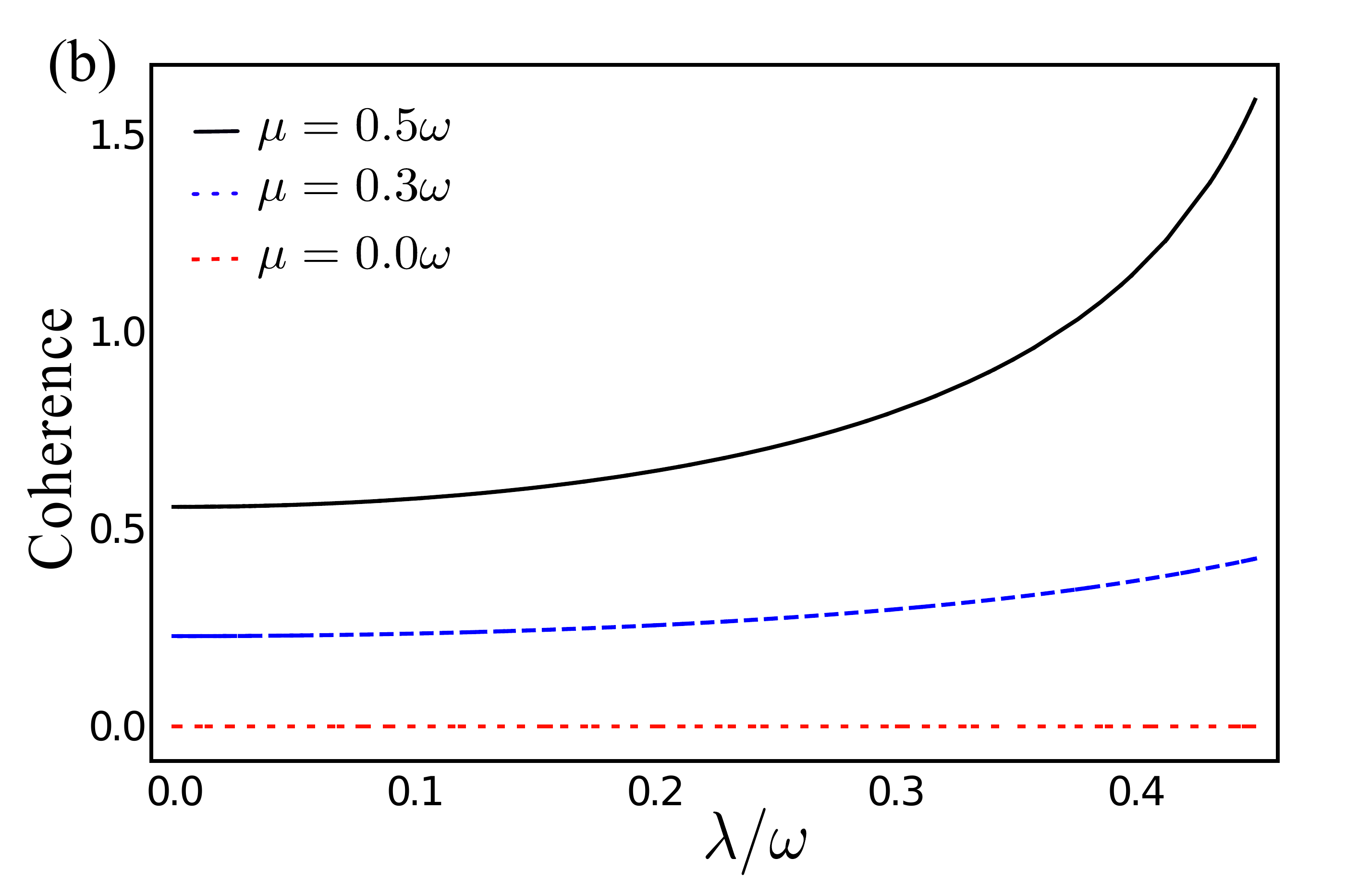}
\caption{Quantum coherence as a function of the intermode coupling parameters for different values of the coupling between the two modes. (a) We consider $\lambda = 0.0\omega$ (black solid line), $\lambda = 0.3\omega$ (blue dashed line), and $\lambda = 0.5\omega$ (red dotted line) . (b) We set $\mu = 0.0\omega$ (red dotted line), $\mu = 0.3\omega$ (blue dashed line), and $\mu = 0.5\omega$ (black solid line).}
\label{freecase}
\end{figure}
 In fact, the results above show that there is coherence production even if the exchange of excitation is null and the coupling is purely squeezing. We can observe that for the zero exchange of excitation degree, the coherence production follows a monotonic behavior as $\mu$ increases, whereas for the zero squeezing degree the coherence production is zero irrespective of the value of $\lambda$. Therefore, although the exchange of excitation contributes to increase the coherence of the system it does not generate coherence.

\section{Coherence of two-coupled bosonic modes: Both interacting with a thermal bath} \label{section02}

\begin{figure}[H]
\centering
\includegraphics[scale=0.40]{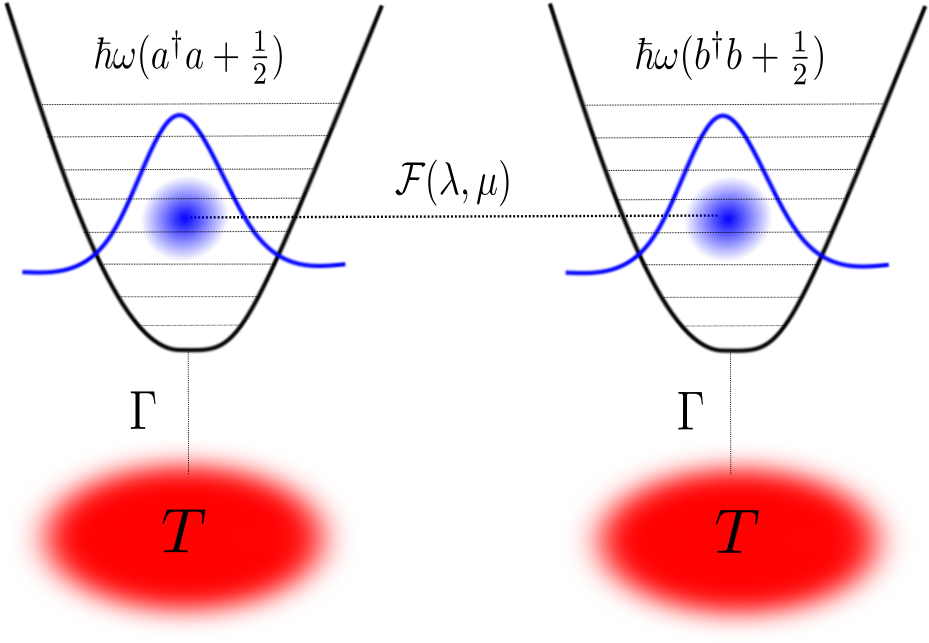}
\caption{Illustration of the system. Two coupled modes both interacting with identical thermal reservoirs with temperature $T$.}
\label{Figure03A}
\end{figure}

We consider here the situation in which both bosonic modes interact with identical Markovian thermal baths with fixed temperature $T$. The interaction between the quantum system and the environment is treated by the  Caldeira- Leggett model where the system-environment coupling is considered sufficiently weak \cite{Joshi2014,BreuerBook}. The new configuration is depicted in Fig. \ref{Figure03A}, where  $\mathcal{F}=\mathcal{F}(\lambda,\mu)$ is the intermode coupling, $\Gamma$ is the system-environment coupling and $T$ is the environment temperature. 

The Hamiltonian that describes the whole system is given by
\begin{equation}
\hat{\mathcal{H}}=\hat{\mathcal{H}}_{S}+\hat{\mathcal{H}}_{B}+\hat{\mathcal{H}}_{SB},
\end{equation}
\begin{equation}
\hat{\mathcal{H}}_{B}=\sum_{n}\omega_{a,n}\hat{a}_{n}^{\dagger}\hat{a}_{n}+\sum_{m}\omega_{b,m}\hat{b}_{m}^{\dagger}\hat{b}_{m},\label{eq:7-1}
\end{equation}
\begin{align}
\hat{\mathcal{H}}_{SB} & =\left(\hat{a}+\hat{a}^{\dagger}\right)\sum_{n}\lambda_{a,n}\left(\hat{a}_{n}+\hat{a}_{n}^{\dagger}\right)\nonumber \\
 & +\left(\hat{b}+\hat{b}^{\dagger}\right)\sum_{m}\lambda_{b,m}\left(\hat{b}_{m}+\hat{b}_{m}^{\dagger}\right),\label{eq:8-1}
\end{align}
where $\hat{\mathcal{H}}_{B}$ and $\hat{\mathcal{H}}_{SB}$ stand for the Hamiltonian of the bath and the Hamiltonian of the system-environment coupling, respectively, and $\hat{\mathcal{H}}_S$ is given by Eq. (\ref{eq:5}).

The dynamics of the two modes independent interacting with Markovian thermal baths is dictated by the following master equation (in the interaction picture)\cite{Joshi2014}
\begin{align}
\frac{d\hat{\rho}_S}{dt} & =-\gamma_{\alpha}\left\{ \left[1+\bar{n}_{B}(\Lambda_{+})\right]\mathcal{L}[\hat{\alpha}](\hat{\rho})+\bar{n}_{B}(\Lambda_{+})\mathcal{L}[\hat{\alpha}^{\dagger}](\hat{\rho})\right\} \nonumber \\
 & -\gamma_{\beta}\left\{ \left[1+\bar{n}_{B}(\Lambda_{-})\right]\mathcal{L}[\hat{\beta}](\hat{\rho})+\bar{n}_{B}(\Lambda_{-})\mathcal{L}[\hat{\beta}^{\dagger}](\hat{\rho})\right\} ,\label{eq:9-1}
\end{align}
where $\mathcal{L}[\hat{\alpha}](\hat{\rho})\equiv\lbrace\hat{\alpha}^{\dagger}\hat{\alpha},\hat{\rho}\rbrace-2\hat{\alpha}\hat{\rho}\hat{\alpha}^{\dagger}$ is the Lindblad operator; $\Lambda_{\pm}=\sqrt{\omega^{2}+\lambda^{2}-\mu^{2}\pm2\lambda\mu}$
are the frequencies of the diagonalized hamiltonian which satisfy the simple relation $\Lambda_{+}=\Lambda_{-}=\sqrt{\omega^{2}+\lambda^{2}}$ if $\mu=0$; $\hat{\alpha}/\hat{\beta}$ is given in Eq. (\ref{eq:50}) and Eq. (\ref{eq:51}), respectively (see Appendix A) and they are the bosonic modes that diagonalize the Hamiltonian in Eq. (\ref{eq:5}); $\bar{n}(\Lambda)=\frac{1}{\exp[\Lambda/T]-1}$ is the Bose-Einstein distribution. The constants $\gamma_{\alpha/\beta}$ are related with the intensity of system-environment coupling. One can show that the master equation above is of nonlocal Lindblad form in the individual mode operators \cite{Joshi2014}. 

\begin{figure}[H]
\centering
\includegraphics[scale=0.35]{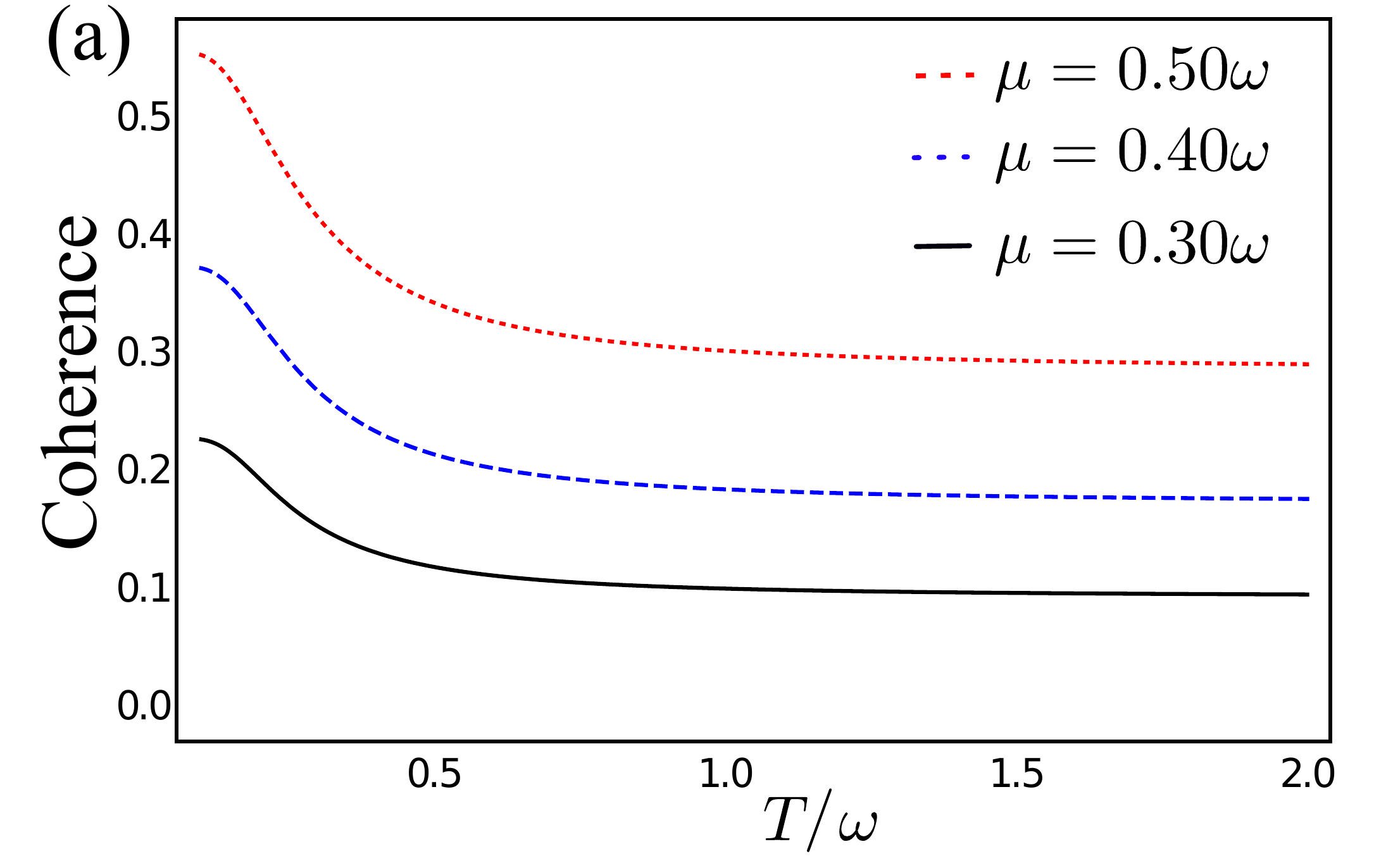}
\includegraphics[scale=0.35]{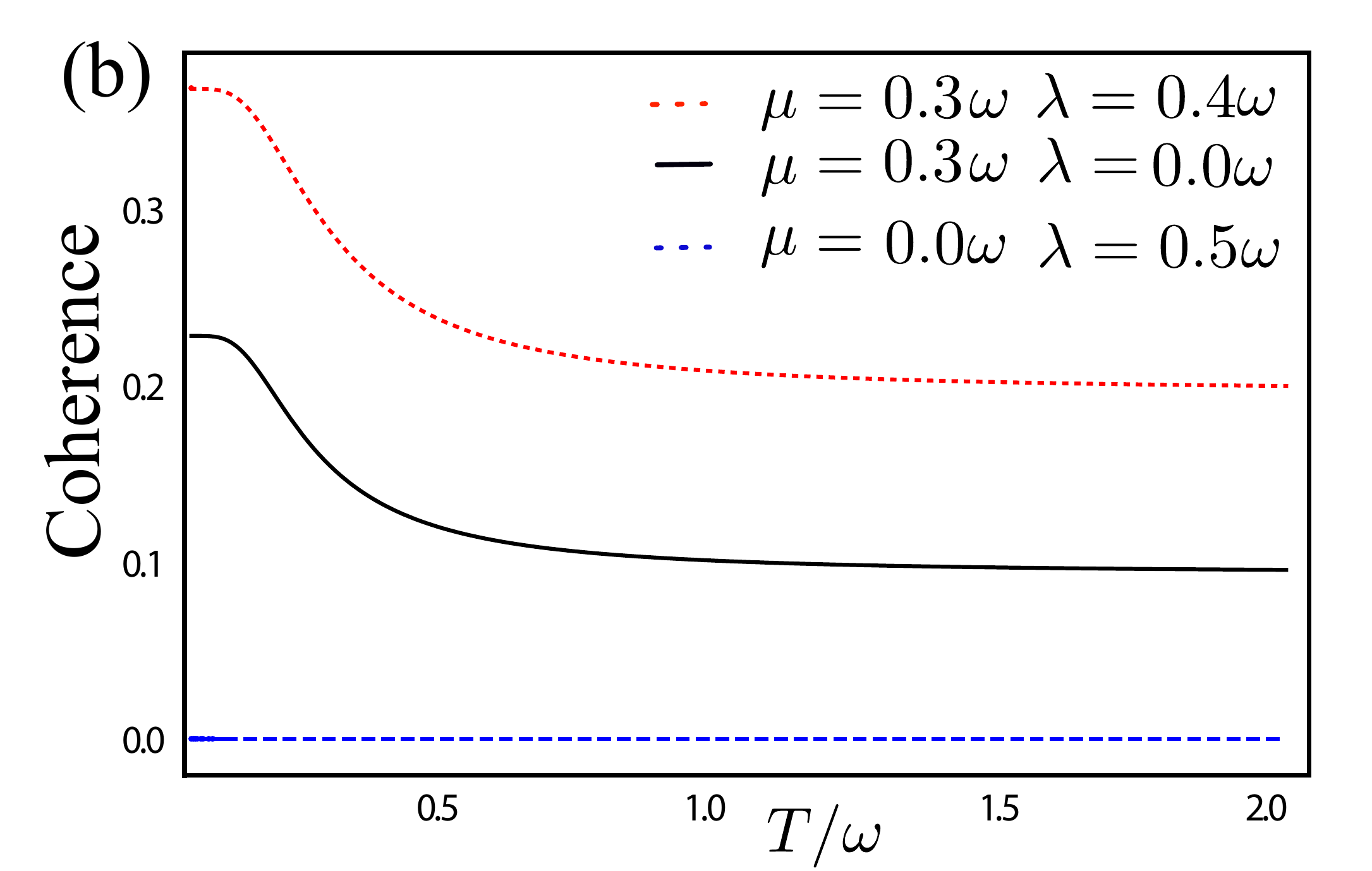}
\caption{Coherence as a function of equilibrium temperature $T$. (a) We consider $\lambda=0.0\omega$ and three values for the squeezing parameter: $\mu = 0.5\omega$ (red dotted line), $\mu = 0.4\omega$ (blue dashed line), and $\mu = 0.3\omega$ (black solid line). (b) we change both parameters: $\mu = 0.3\omega$ and $\lambda = 0.4\omega$ (red dotted line), $\mu = 0.3\omega$ and $\lambda = 0.0\omega$ (black solid line), and $\mu = 0.0\omega$ and $\lambda = 0.5\omega$ (blue dashed line).}
\label{coherencebothres1}
\end{figure}

After solving the above master equation we obtain the following expression for the steady-state, which is a tensor product of two thermal states \cite{new02}
\begin{equation}
\hat{\rho}_S=\mathcal{Z}_{+}\exp\left\{ -\frac{\Lambda_{+}}{T}\hat{\alpha}^{\dagger}\hat{\alpha}\right\} \otimes \mathcal{Z}_{-}\exp\left\{ -\frac{\Lambda_{-}}{T}\hat{\beta}^{\dagger}\hat{\beta}\right\} ,\label{eq:10-1}
\end{equation}
where $\mathcal{Z}_{\pm}=1-\exp\left[-\frac{\Lambda_{\pm}}{T}\right]$ is the associated partition function. We stress for the fact that Eq. (\ref{eq:10-1}) is a tensor product of two thermal states in the effective modes $\left(\hat{\alpha}^\dagger\hat{\alpha}, \hat{\beta}^\dagger\hat{\beta}\right)$ but not for the original modes $\left(\hat{a}^\dagger\hat{a}, \hat{b}^\dagger\hat{b}\right)$. From this state, we calculate the Wigner function (see Appendix B) and the elements of the corresponding covariance matrix, 
\begin{equation}
\left\langle \hat{x}_{a}^{2}\right\rangle =\left\langle \hat{x}_{b}^{2}\right\rangle =\frac{1}{16}\left(\frac{1}{\kappa_{1}\tanh{(\frac{\Lambda_{+}}{2T}})}+\frac{1}{\kappa_{2}\tanh{(\frac{\Lambda_{-}}{2T}})}\right),\label{eq:12-1}
\end{equation}
\begin{equation}
\left\langle \hat{x}_{a}\hat{x}_{b}\right\rangle =\frac{1}{16}\left(\frac{1}{\kappa_{1}\tanh{(\frac{\Lambda_{+}}{2T}})}-\frac{1}{\kappa_{2}\tanh{(\frac{\Lambda_{-}}{2T}})}\right),\label{eq:13-1}
\end{equation}
\begin{equation}
\left\langle \hat{p}_{a}^{2}\right\rangle =\left\langle \hat{p}_{b}^{2}\right\rangle =\left(\frac{\kappa_{1}}{\tanh{(\frac{\Lambda_{+}}{2T}})}+\frac{\kappa_{2}}{\tanh{(\frac{\Lambda_{-}}{2T}})}\right),\label{eq:14-1}
\end{equation}
\begin{equation} 
\left\langle \hat{p}_{a}\hat{p}_{b}\right\rangle =\left(\frac{\kappa_{1}}{\tanh{(\frac{\Lambda_{+}}{2T}})}-\frac{\kappa_{2}}{\tanh{(\frac{\Lambda_{-}}{2T}})}\right).\label{eq:15-1}
\end{equation}
We calculate the coherence of the system of bosonic modes interacting with the environment from the covariance matrix in Eq. (\ref{eq:18}) and the elements from Eq. (\ref{eq:12-1}) to Eq. (\ref{eq:15-1}). In Fig. \ref{coherencebothres1} we show the behavior of the coherence as a function of equilibrium temperature $T$. In Fig. \ref{coherencebothres1}-(a) we consider a purely squeezing intermode coupling, i.e., $\lambda = 0.0\omega$ and three values for the squeezing parameter: $\mu = 0.5\omega$ (red dotted line), $\mu = 0.4\omega$ (blue dashed line), and $\mu = 0.3\omega$ (black solid line). In Fig. \ref{coherencebothres1}-(b) we change both parameters: $\mu = 0.3\omega$ and $\lambda = 0.4\omega$ (red dotted line), $\mu = 0.3\omega$ and $\lambda = 0.0\omega$ (black solid line), and $\mu = 0.0\omega$ and $\lambda = 0.5\omega$ (blue dashed line). 
We observe that for a purely squeezing intermode coupling the coherence is degraded by the environment temperature. However, we still observe that more coherence is associated with more squeezing. Another feature that can be observed is that the coherence does not go to zero as a function of the environment temperature, but to a constant value that depends on the value of the squeezing. This behavior is different from the sudden death behavior obtained from the entanglement in Ref. \cite{Joshi2014}. When we change both parameters we can observe by the blue dashed line in Fig. \ref{coherencebothres1}-(b) that again there is no coherence for zero squeezing. We highlight that this behavior is valid for any value of $\lambda$ provided that $\mu = 0$. We can also observe that for the same value of the squeezing the effect of the temperature is less destructive if the exchange of excitation is present. This last effect is a consequence of the nonlocal Lindblad terms which contribute to mixing some excited states with the ground state and thus the steady-state is not separable in the ground state of each individual mode. Then, the presence of the exchange of excitation, as well as the energy conservation, will partially protect the quantum system of the temperature effect.

The constant value attained by the coherence in Fig. 4, as a function of the intermode coupling parameters $\mathcal{C}_{\rho}(\mu,\lambda)$, was calculated in Appendix C by taking the limit $T\rightarrow\infty$. 

\section{Coherence of two-coupled bosonic modes: Only one interacting with a thermal bath}\label{section03}

In order to go further in our investigation, we now assume that only one mode is interacting with the thermal bath and study the coherence dynamics for an initial pure Gaussian state. We denote this particular mode as our system of interest, whereas the other mode is designated as the ancilla. To simplify our study we consider that the system is free from the environment at the time $t=0$ and the initial state of the system+ancilla is pure. We can also consider the environment effect at $t=0$ and evolve the corresponding initial mixed state but the treatment is more difficult to deal with. We will see that depending on the choice of parameters, the combination of the Markovian thermal bath and the ancilla works effectively as a non-Markovian environment, having as a signature the non-monotonic behavior of an appropriated quantum information quantity \cite{Rivas2010, Breuer2009, Breuer2016, Rajagopal2010, Santos2021}. Writing the position and momentum operators as $\hat{x}_i= \sqrt{1/(2\omega)}(\hat{c}_i^\dagger + \hat{c}_i)$ and  $\hat{p}_i=i\sqrt{\omega/2}(\hat{c}_i^\dagger - \hat{c}_i)$, with $\hat{c}_i = (\hat{a}, \hat{b})$, it is possible to obtain the associated Wigner function for the system. We then proceed to consider the bosonic mode coupled to the thermal environment as the system of interest, i.e., $\mathcal{W}_{sys}\equiv \mathcal{W}_{a}(x_{a},p_{a})$. This setup allows us to investigate how the thermalization dynamics of a bosonic mode is impacted by its surrounding, the role played by the second mode. 

\begin{figure}[H]
\centering
\includegraphics[scale=0.30]{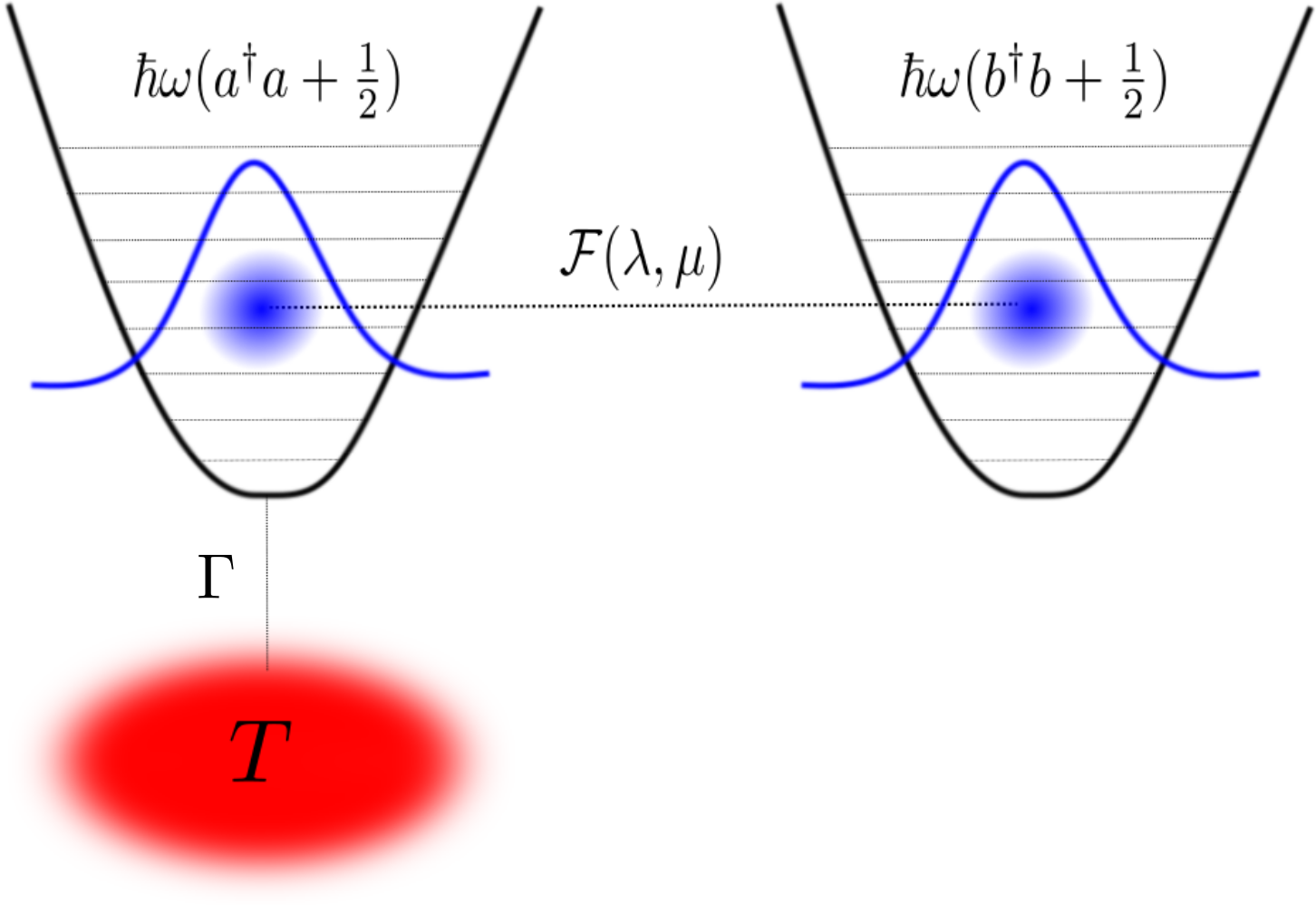}
\caption{Illustration of the system. Only one of the modes is coupled with the environment. We denote by $\left(a^\dagger, a\right)$ and $\left(b^\dagger, b\right)$ the system and ancilla variables, respectively. }
\label{lastsetup}
\end{figure}

Figure \ref{lastsetup} illustrates the present scenario. The coupling between the system and the thermal environment is mediated by the coupling $\Gamma$, while the constant $\mathcal{F}=\mathcal{F}(\lambda,\mu)$ controls the coupling between the system and the ancilla. We stress that the interaction with the thermal bath is assumed to be sufficiently weak in order to satisfy the Born-Markov approximation \cite{BreuerBook}.

In order to calculate the coherence of the system $\mathcal{W}_{sys}\equiv \mathcal{W}_{a}(x_{a},p_{a})$, we follow the steps \cite{new08}: 
(I) we construct the equations of motion using the Heisenberg equation for $x_\pm$ and $p_\pm$, which correspond to the decoupled modes $\hat{\alpha}$ and $\hat{\beta}$, respectively, in Eq. (\ref{eq:14}) such that $x_+ \propto (\hat{\alpha}^\dagger + \hat{\alpha})$ and  $p_+ \propto (\hat{\alpha}^\dagger - \hat{\alpha})$ and the same is valid for $x_-$ and $p_-$ relative to $\hat{\beta}$ and $\hat{\beta}^\dagger$,
\begin{align}
 x_{\pm_{(0)}}&=\cos{(\Lambda_{\pm}t)}x_{\pm}-\frac{\sin{(\Lambda_{\pm}t)}}{\Lambda_{\pm}}p_{\pm}\\
 p_{\pm_{(0)}}&=\Lambda_{\pm}\sin{(\Lambda_{\pm}t)}x_{\pm}+\cos{(\Lambda_{\pm}t)}p_{\pm},
\end{align}
where $x_{\pm_{(0)}}$ and $p_{\pm_{(0)}}$ are the position and momentum values at $t=0$ and $x_{\pm}$, and $p_{\pm}$ are the analog values at time $t$ \cite{new08}. 

(II) we replace the results above in the pure two-mode Gaussian state
\begin{align}
\mathcal{W}(x_{i},p_{i};t) &=\frac{1}{\pi^{2}}\exp{\left[-\left(x_{+_{(0)}}\right)^{2}-\left(p_{+_{(0)}}\right)^{2}\right]} \nonumber \\
&\times\exp{\left[-\left(x_{-_{(0)}}\right)^{2}-\left(p_{-_{(0)}}\right)^{2}\right]},
\end{align}

(III) Finally, we take the partial trace of $\mathcal{W}(x_{i},p_{i};t)$ in the coordinates 
$(x_{b},p_{b})$ of the second mode to obtain the Wigner function $\mathcal{W}_{sys}$.
Then, we analyze the dissipative dynamics of the main system in contact with a Markovian environment. Physically, when we restrain our attention only to the main system $(x_{a},p_{a})$, we are assuming that the coupling between the thermal environment and the degrees of freedom of the ancilla $(x_{b},p_{b})$ is sufficiently weak such that it does not yield any consequence on the evolution of the principal system. In turn, since the main system is Gaussian we can encode the thermalization process with the Markovian bath completely in terms of the first moments and covariance matrix during the dissipative process as has been done in the references \cite{Santos2021A,Serafini,Giovanneti}.

\begin{figure}[H]
\centering
	\includegraphics[scale=0.4]{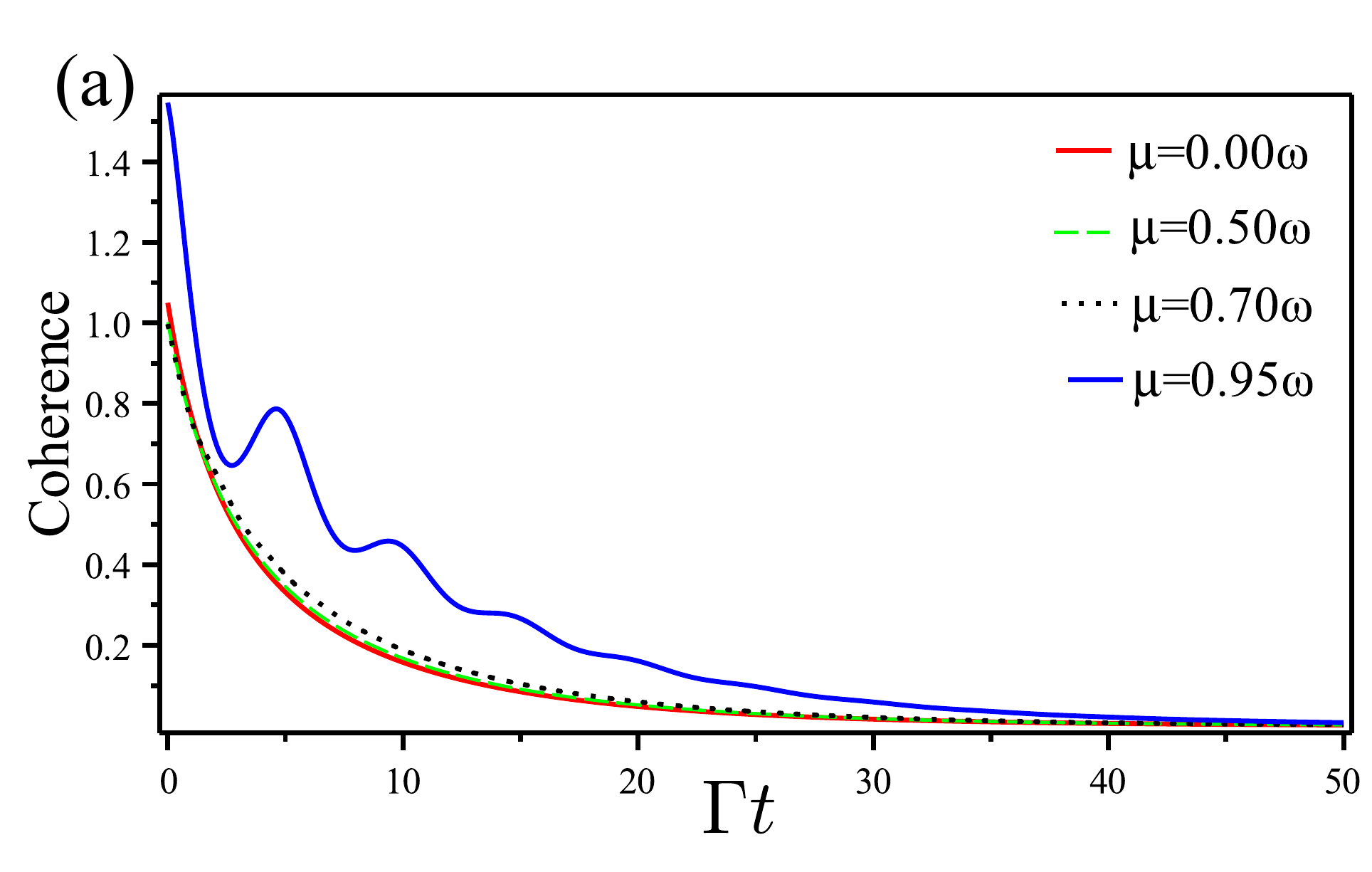}
	\includegraphics[scale=0.4]{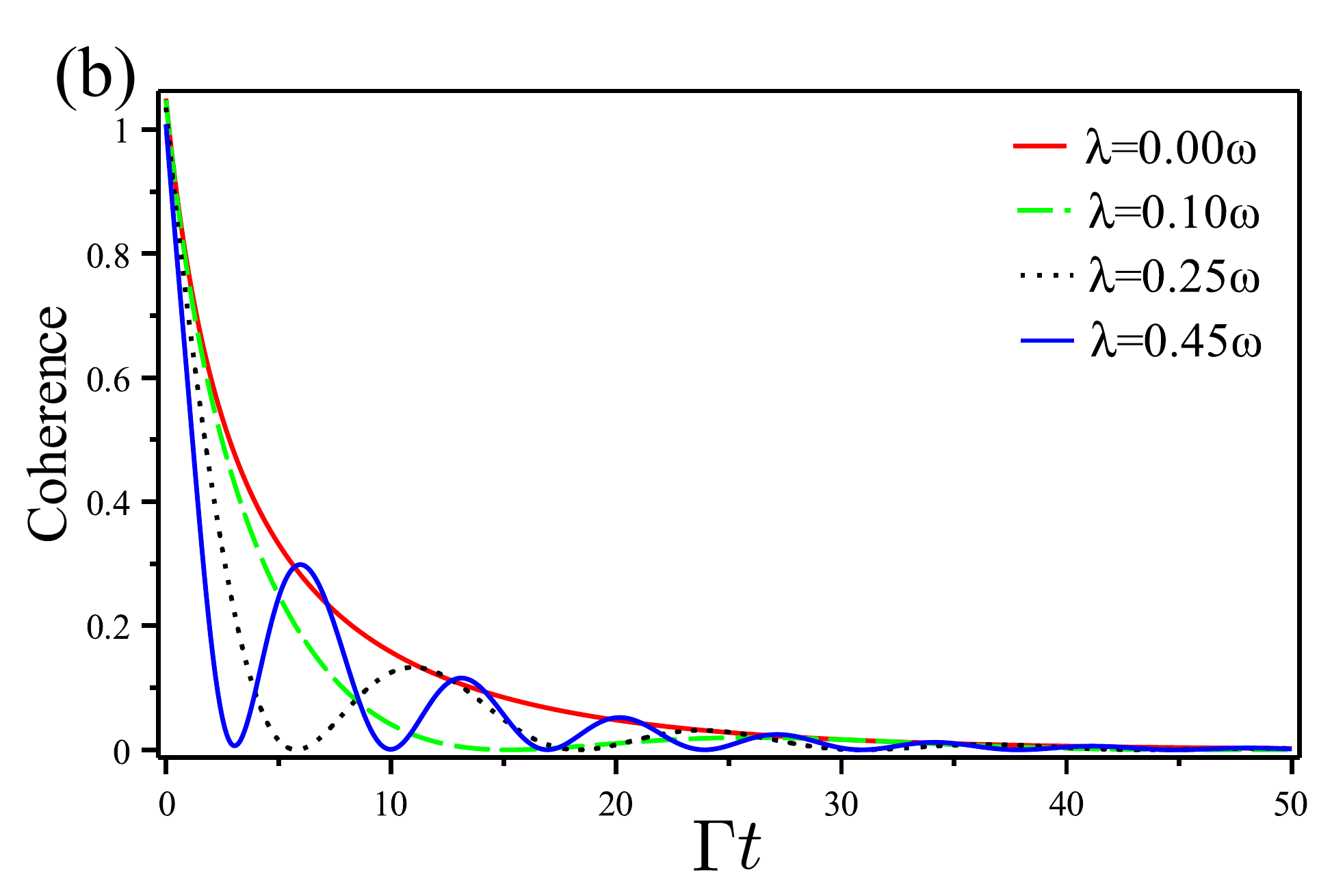}
	\caption{Quantum coherence for the mode coupled to the environment as a function of the time for different values of the coupling between the two modes. In Fig. 6-(a) we set $\lambda=0$ and change the squeezing parameter and in Fig. 6-(b) we set $\mu=0$ and change the value of the exchange of excitation parameter. We have set the initial coordinates to be $(1,1,1,1)$.}
\label{coherencelastsetup}
\end{figure}

In Fig. \ref{coherencelastsetup} we show the quantum coherence for the system of interest coupled to the environment as a function of the time for different values of the coupling between the two modes. In Fig. \ref{coherencelastsetup}-(a) we fix $\lambda=0$ and change the squeezing parameter. We observe that the system already begins the dissipative dynamics with an initial amount of coherence because the initial state is displaced (we set the initial coordinates to be $(1,1,1,1)$). This is why the initial coherence is non-zero even for the zero squeezing parameter (red curve), and the initial coherence increases for other values of $\mu$ (green, black and blue curves). In Fig. \ref{coherencelastsetup}-(b) we have set $\mu=0$ and changed the value of the exchange of excitation parameter. In this last case, it is observed that the initial coherence is the same for all cases because the exchange of excitation parameter does not generate coherence. We note that, in both cases, when the ancilla effect is present, in general, the monotonic behavior of the coherence is not achieved. 

\begin{figure}[H]
\centering
	\includegraphics[scale=0.4]{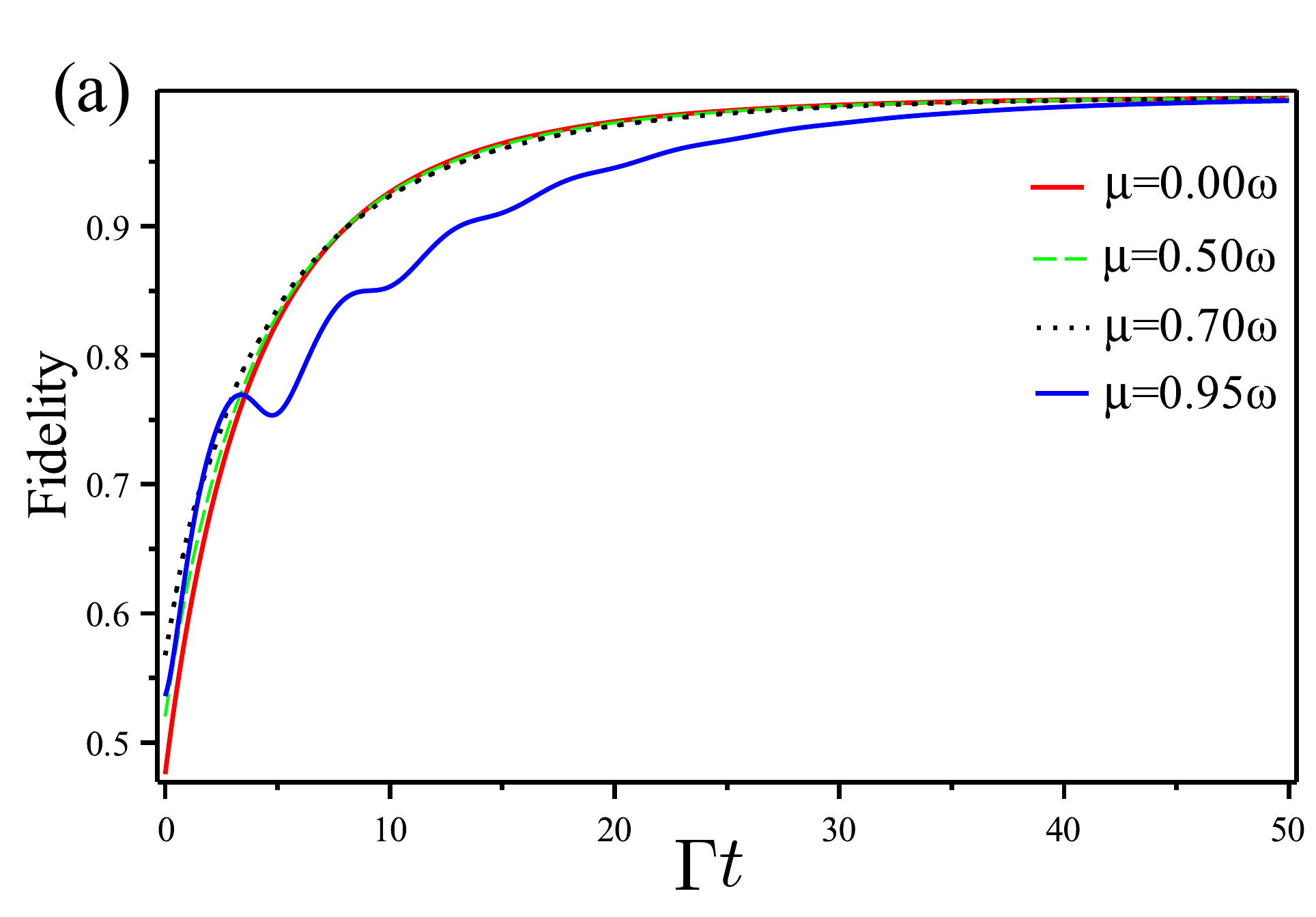}
	\includegraphics[scale=0.4]{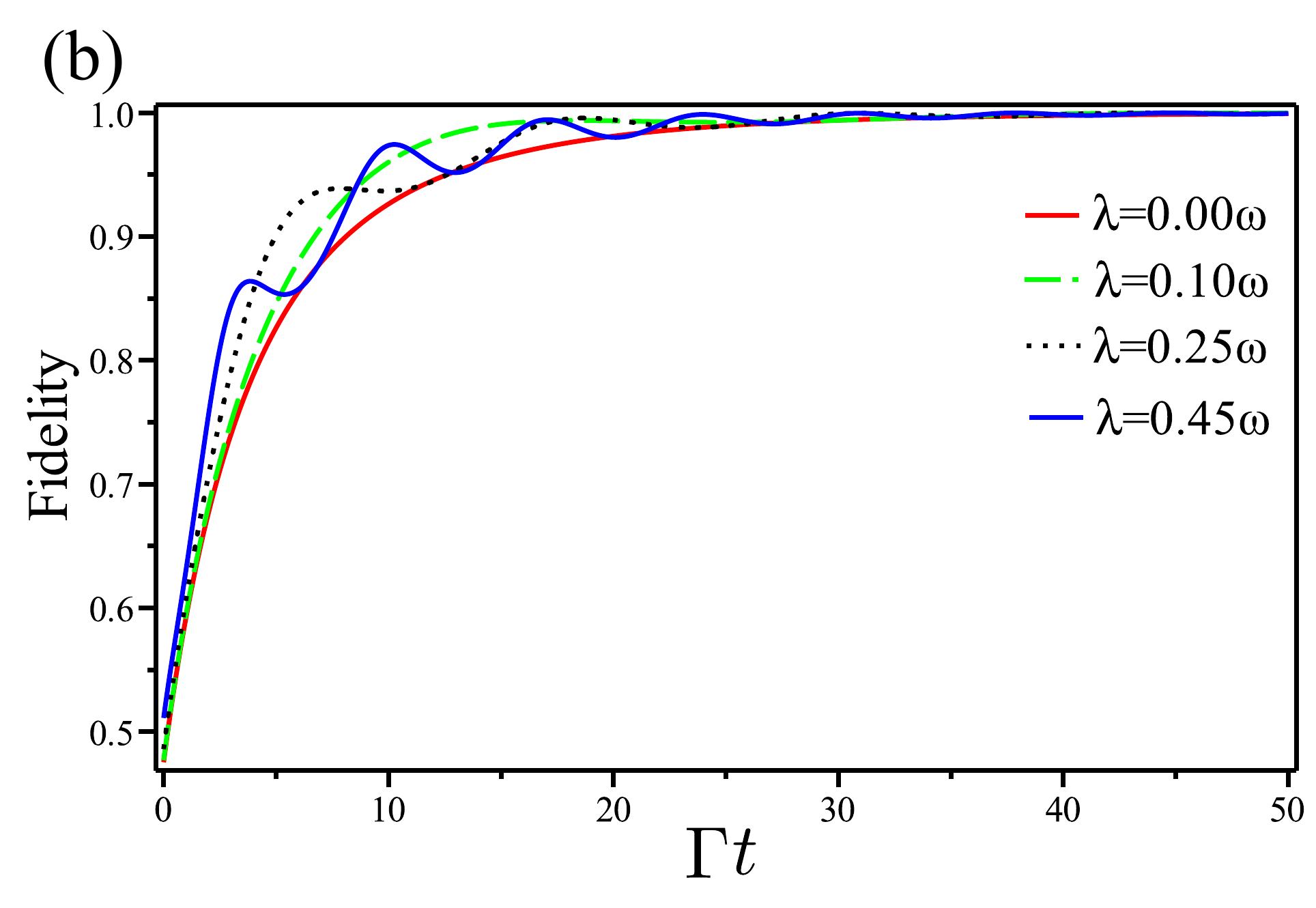}
	\caption{Fidelity for the mode coupled to the environment as a function of the time for different values of the coupling between the two modes. In Fig. 7-(a) we set $\lambda=0$ and change the squeezing parameter and in Fig. 6-(b) we set $\mu=0$ and change the value of the exchange of excitation parameter.  We have set the initial coordinates to be $(1,1,1,1)$.}
\label{fidelity}
\end{figure}

In Fig. \ref{fidelity}  we show the fidelity to compare the state of the main system which is coupled to the environment and the asymptotic state as a function of the time for different values of the intermode coupling between the system and ancilla. In Fig. 7-(a) we have set $\lambda=0$ and changed the squeezing parameter and in Fig. 7-(b) we have set $\mu=0$ and changed the value of the exchange of excitation parameter. When the two modes are decoupled, i.e., ($\mathcal{F}(\lambda,\mu)=0$), the dissipative dynamics of the fidelity has a monotonic increase which is also characteristic of Markovian dynamics, since we are comparing with the asymptotic thermal state. For $\mathcal{F}(\lambda,\mu)\neq0$, the influence of the squeezing parameter $\mu$ on the fidelity is decreasing it with a small oscillation in comparison with the decoupled case as we can observe in Fig. 7-(a). On the other hand, the excitation parameter $\lambda$ produces considerable oscillations in the fidelity while leaving it bigger than the decoupled case as we can observe in Fig. 7-(b). 
When an oscillatory behavior on the coherence and/or fidelity is produced by the interaction with the system and environment we have a non-Markovian effect \cite{Santos2021A, Rajagopal2010, new04, new05} . Here, the oscillatory behavior is produced by the coupling $\mathcal{F}(\lambda,\mu)$ between the modes. Therefore, the intermode coupling can produce a \textit{Non-Markovian-Like effect} when the system is weakly coupled with the environment. A similar behavior has been observed in Ref.  \cite{Santos2021A} for two-coupled bosonic modes with a different interaction and in Ref. \cite{new07}  when the system and the ancilla are two-level systems.

\begin{figure}[h]
\centering
	\includegraphics[scale=0.4]{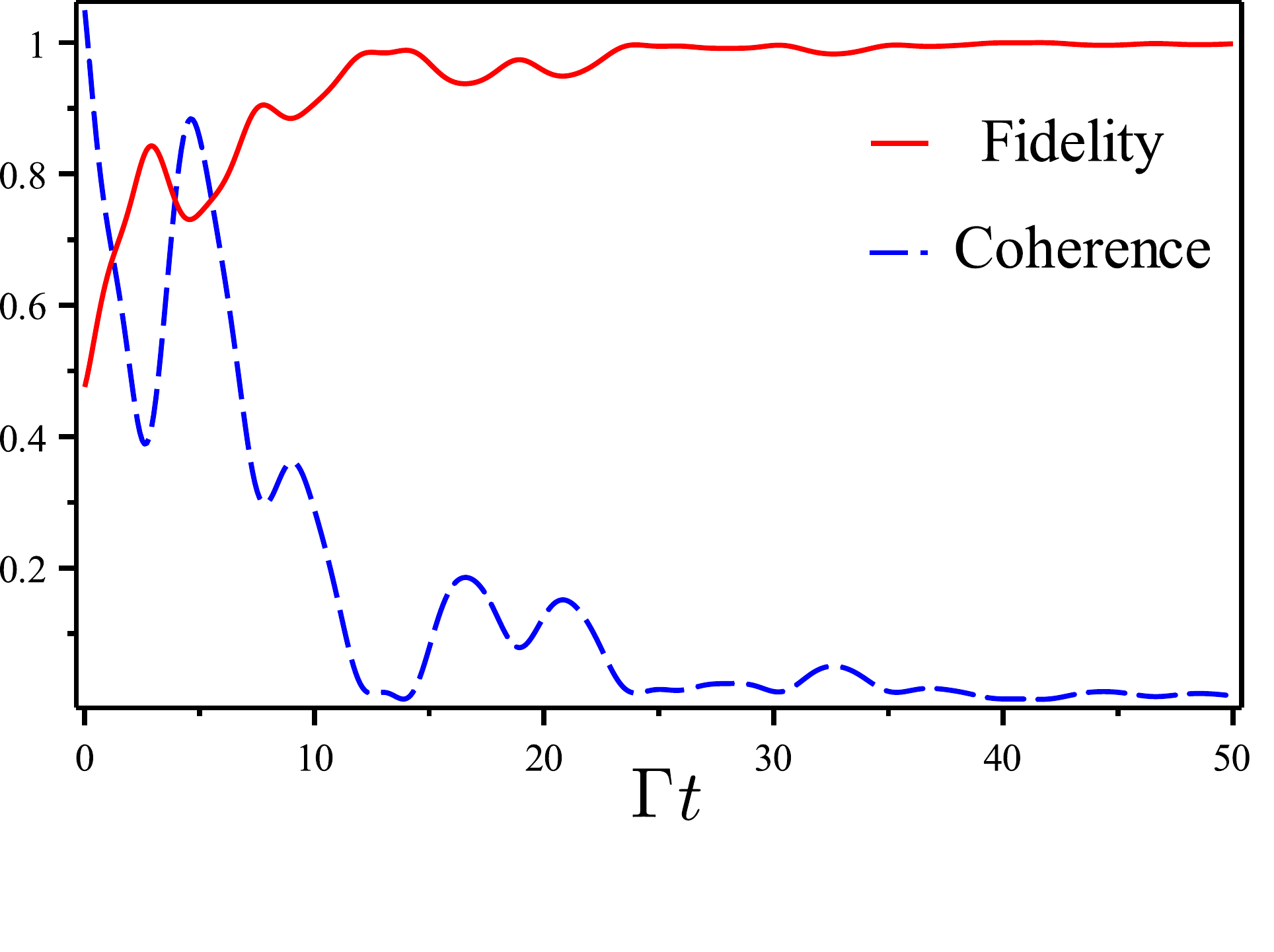}
	\caption{Dissipative dynamics of the coherence and fidelity for the mode coupled with the bath and for the parameters $\mu=0.50\omega$ and $\lambda=0.45\omega$ for the coupling between the modes.  We have set the initial coordinates to be $(1,1,1,1)$.}
	\label{disscoherence}
\end{figure}

In Fig. \ref{disscoherence} we compare the dissipative dynamics of the coherence and fidelity for the mode coupled with the bath and for the parameters $\mu=0.50\omega$ and $\lambda=0.45\omega$ for the coupling between the modes. As we can observe both quantities oscillate as a function of time and when one of them has a maximum the other one has a minimum. For long interaction time, the coherence is null while the fidelity is 1. 

\section{Conclusion} \label{Conclusion}
We have studied the effect of the intermode coupling in the quantum coherence of a system formed by two coupled bosonic modes when they are free from the environment, when each mode interacts with identical Markovian baths, and when one of the modes interacts with a Markovian bath while the other one remains free. We separated the interaction Hamiltonian in weak and strong coupling effects with respect with the squeezing part of the coupling. For weak coupling, the RWA applies which means negligible squeezing and for strong coupling, it does not apply. Then, we studied the behavior of coherence and we observed that to produce coherence the coupling has to present a squeezing term which means that coherence is produced by strong intermode coupling. On the other hand, a pure exchange of excitation intermode coupling does not produce coherence even if this coupling is strong. In fact, we observed that the exchange of excitation contributes to increasing the coherence produced by the squeezing interaction. 

When the two modes are coupled with identical Markovian baths we found that the coherence decreases with the bath temperature saturating to a nonzero constant value which is intermode coupling dependent. As in the case free from the environment, the coherence is associated with the squeezing interaction and the exchange of excitation contributes to reducing the effect of the temperature. We observed that the coherence has a behavior similar to the entanglement at $T=0$ but it does not present a sudden death effect when the temperature grows. Finally, we studied the coherence and the fidelity when only one of the two coupled modes interacts with a Markovian bath. We considered that at $t=0$ the modes are free from the environment, i.e., the initial state is pure. We observed that the intermode parameters produce oscillations in the coherence and the fidelity similar to a non-Markovian behavior induced by a strong interaction with a thermal bath. We obtained that the squeezing parameter contributes slightly to the oscillations and strongly with the increase of the coherence and decrease of the fidelity. Unlike, the exchange of excitation parameter contributes strongly to the oscillations of these quantities. 

Finally, our results can be straightforwardly generalized for a bosonic modes chain, allowing us to investigate how the coherence due to the intermode couplings impact quantities as energy transfer. Furthermore, the effects presented by the squeezing and the exchange of excitation during the dynamics indicate that they could be employed to protect the system from decoherence when in contact with the environment.

\section*{Acknowledgments}
The authors would like to thank CAPES and CNPq-Brazil for their financial support. I.G.P. acknowledges Grant No. 307942/2019-8 from CNPq. J.F.S. and C.H.S.V. acknowledge CAPES (Brazil) for support. J. F. G. Santos acknowledges S\~ao Paulo, Research Grant No. 2019/04184-5, and NSFC (China) under Grant No. 12050410258. 

\section{Appendix A: Ground state and quantum coherence}

In this section, we diagonalize the Hamiltonian in Eq.(\ref{eq:5})
and find the covariance matrix for the ground state. We can start
by rewriting it in the following form 
\begin{equation}
\hat{\mathcal{H}}=\omega\hat{a}^{\dagger}\hat{a}+\omega\hat{b}^{\dagger}\hat{b}+\lambda\left(\hat{a}^{\dagger}+\hat{a}\right)\left(\hat{b}^{\dagger}+\hat{b}\right)+(\mu-\lambda)\left(\hat{a}^{\dagger}\hat{b}^{\dagger}+\hat{a}\hat{b}\right).\label{eq:6}
\end{equation}

To decouple the Hamiltonian in Eq.(\ref{eq:6}) we use the center of
mass and relative coordinates between the modes. More precisely, we
used the transformations 
\begin{equation}
\hat{R}=\frac{1}{2}(\hat{x}_{a}+\hat{x}_{b}),\;\hat{r}=(\hat{x}_{a}-\hat{x}_{b}),\label{eq:7}
\end{equation}
the same is valid for momentum coordinates. Defining creation and
annihilation operators for the center of mass $\hat{A}^{\dagger}$($\hat{A}$)
and relative coordinates $\hat{B}^{\dagger}$($\hat{B}$), we find a relationship
between the first and last operators 
\begin{equation}
\left(\begin{array}{c}
\hat{A}\\
\hat{A}^{\dagger}\\
\hat{B}\\
\hat{B}^{\dagger}
\end{array}\right)=\mathbf{M}\cdot\left(\begin{array}{c}
\hat{a}\\
\hat{a}^{\dagger}\\
\hat{b}\\
\hat{b}^{\dagger}
\end{array}\right),\label{eq:8}
\end{equation}
where the transformation matrix is given by
\begin{equation}
\mathbf{M}=\frac{1}{\sqrt{8}}\left(\begin{array}{lccr}
c_{1} & c_{2} & c_{1} & c_{2}\\
c_{2} & c_{1} & c_{2} & c_{1}\\
c_{3} & c_{4} & -c_{3} & -c_{4}\\
c_{4} & c_{3} & -c_{4} & -c_{3}
\end{array}\right)\ \textrm{;}\ \left\{ \begin{array}{lll}
c_{1}={\displaystyle \sqrt{\frac{\omega_{+}}{\omega}}+{\displaystyle \sqrt{\frac{\omega}{\omega_{+}}}}}\\
c_{2}={\displaystyle \sqrt{\frac{\omega_{+}}{\omega}}-{\displaystyle \sqrt{\frac{\omega}{\omega_{+}}}}}\\
c_{3}={\displaystyle \sqrt{\frac{\omega_{-}}{\omega}}+{\displaystyle \sqrt{\frac{\omega}{\omega_{-}}}}}\\
c_{4}={\displaystyle \sqrt{\frac{\omega_{-}}{\omega}}-{\displaystyle \sqrt{\frac{\omega}{\omega_{-}}}}}
\end{array}\right.,\label{eq:9}
\end{equation}
being the frequencies of the virtual center of mass and relative coordinate,
$\omega_{\pm}\equiv\sqrt{\omega^{2}\pm2\lambda\omega}$ holding the
condition $2\lambda<\omega$ since the eigenenergies of the system need to be positive real-valued in order to obtain stabe solutions \cite{Wang2020}.

Writing the Hamiltonian Eq.(\ref{eq:6}) in terms of the new operators
as follows 
\begin{eqnarray}
\hat{\mathcal{H}}&=\vartheta_{+}\hat{A}^{\dagger}\hat{A}+\zeta_{+}\left(\hat{A}^{\dagger}\hat{A}^{\dagger}+\hat{A}\hat{A}\right)\nonumber\\&+\vartheta_{-}\hat{B}^{\dagger}\hat{B}+\zeta_{-}\left(\hat{B}^{\dagger}\hat{B}^{\dagger}+\hat{B}\hat{B}\right),\label{eq:10}
\end{eqnarray}
where 
\begin{equation}
\vartheta_{\pm}=\frac{\lambda^{2}+\omega^{2}-\lambda\mu\pm2\lambda\omega}{\sqrt{\omega(\omega\pm2\lambda)}}\;,\zeta_{\pm}=\frac{(\mu-\lambda)(\lambda\pm\omega)}{2\sqrt{\omega(\omega\pm2\lambda)}}.\label{eq:11}
\end{equation}
The Hamiltonian in Eq.(\ref{eq:10}) does not contain any mixed
terms between the new operators $\hat{A}$ and $\hat{B}$. We can split it into
two parts containing either $\hat{A}$ and $\hat{B}$. Moreover, the terms proportional to $\zeta_{+}$ and  $\zeta_{-}$ are responsible for generating squeezing on an initial state. The final diagonalization
of the Hamiltonian operator is provided by the transformation
\begin{equation}
\left(\begin{array}{c}
\hat{\alpha}\\
\hat{\alpha}^{\dagger}\\
\hat{\beta}\\
\hat{\beta}^{\dagger}
\end{array}\right)=\left(\begin{array}{cccc}
u_{a} & v_{a} & 0 & 0\\
v_{a} & u_{a} & 0 & 0\\
0 & 0 & u_{b} & v_{b}\\
0 & 0 & v_{b} & u_{b}
\end{array}\right)\cdot\left(\begin{array}{c}
\hat{A}\\
\hat{A}^{\dagger}\\
\hat{B}\\
\hat{B}^{\dagger}
\end{array}\right),\label{eq:12}
\end{equation}
with $u_{a,b}=\cosh(r_{a,b})$, $v_{a,b}=\sinh(r_{a,b})$ and
\begin{equation}
r_{a/b}=-\frac{1}{2}\textrm{arctanh}{\left(\frac{(\lambda-\mu)(\lambda\pm\omega)}{\lambda^{2}-\lambda\mu\pm2\lambda\omega+\omega^{2}}\right).}\label{eq:13}
\end{equation}

In terms of $\hat{a} (\hat{a}^{\dagger})$ and $\hat{b} (\hat{b}^{\dagger})$, we have
\begin{align}
\hat{\alpha} &= \frac{1}{\sqrt{8}}[(c_{1}\cosh(r_{a})+c_{2}\sinh(r_{a}))(\hat{a}+\hat{b})\nonumber \\
&+ (c_{2}\cosh(r_{a})+c_{1}\sinh(r_{a}))(\hat{a}^{\dagger}+\hat{b}^{\dagger})], \label{eq:50}
\end{align}
\begin{align}
\hat{\beta} &= \frac{1}{\sqrt{8}}[(c_{3}\cosh(r_{b})+c_{4}\sinh(r_{b}))(\hat{a}-\hat{b})\nonumber \\
&+ (c_{4}\cosh(r_{b})+c_{3}\sinh(r_{b}))(\hat{a}^{\dagger}-\hat{b}^{\dagger})]. \label{eq:51}
\end{align}

Therefore, we find that
\begin{equation}
\hat{\mathcal{H}}=\Lambda_{+}\hat{\alpha}^{\dagger}\hat{\alpha}+\Lambda_{-}\hat{\beta}^{\dagger}\hat{\beta}+\textrm{const},\label{eq:14}
\end{equation}
with is simply the Hamiltonian of two uncoupled bosonic modes with
frequencies  $\Lambda_{\pm}=\sqrt{\omega^{2}+\lambda^{2}-\mu^{2}\pm2\lambda\mu}$.
As consequence, the ground state is therefore the product of two Gaussian
functions
\begin{equation}
\Psi(x_{+},x_{-})=\left(\frac{\Lambda_{+}}{\pi}\right)^{\frac{1}{4}}e^{-\frac{\Lambda_{+}x_{+}^{2}}{2}}\left(\frac{\Lambda_{-}}{\pi}\right)^{\frac{1}{4}}e^{-\frac{\Lambda_{-}x_{-}^{2}}{2}}.\label{eq:15}
\end{equation}
We can rewrite the coordinates $x_{\pm}$ back to the original
variables $x_{a/b}$ yielding,
\begin{equation}
 x_{+} = \sqrt{\frac{\omega_{+}}{2\Lambda_{+}}}(v_{a} + u_{a})(x_{a} + x_{b}),
\end{equation}
\begin{equation}
 x_{-} = \sqrt{\frac{\omega_{-}}{2\Lambda_{-}}}(v_{b} + u_{b})(x_{a} - x_{b}).
\end{equation}

Finally, we obtain the ground state 
\begin{align}
\Psi(x_{a},x_{b}) & =\frac{2\cdot\sqrt[4]{\kappa_{1}\kappa_{2}}}{\sqrt{\pi}}\cdot\exp{\left[-\kappa_{1}(x_{a}+x_{b})^{2}\right]}\times\nonumber \\
 & \exp{\left[-\kappa_{2}(x_{a}-x_{b})^{2}\right]},\label{eq:16}
\end{align}
with 
\begin{equation}
\kappa_{1}=\frac{\omega\sqrt{\omega+\lambda+\mu}}{4\cdot\sqrt{\omega+\lambda-\mu}} \quad \textrm{;} \quad \kappa_{2}=\frac{\omega\sqrt{\omega-\lambda-\mu}}{4\cdot\sqrt{\omega-\lambda+\mu}}.\label{eq:17}
\end{equation}
Since that the wave function must be normalized the parameters $\kappa_{1/2}$
need to be real. Hence, are defined some conditions: $\lambda, \mu\geq0$,
$\omega>2\lambda$ and $\omega>\lambda+\mu$.

\textbf{A.1 Coherence for two modes Gaussian state}

In doing so, we can use the Eq.(\ref{eq:3}) to show explicitly the
quantum coherence for the ground state of two coupled bosonic modes
as following
\begin{align}
\mathcal{C}[\rho(\sigma,\vec{d})] & =\frac{(\nu_{1}-1)}{2}\ln\left(\frac{\nu_{1}-1}{2}\right)-\frac{(\nu_{1}+1)}{2}\ln\left(\frac{\nu_{1}+1}{2}\right)\nonumber \\
 & +\left(\bar{\varepsilon}_{1}+1\right)\ln\left(\bar{\varepsilon}_{1}+1\right)-\left(\bar{\varepsilon}_{1}\right)\ln\left(\bar{\varepsilon}_{1}\right)\nonumber \\
 & +\frac{(\nu_{2}-1)}{2}\ln\left(\frac{\nu_{2}-1}{2}\right)-\frac{(\nu_{2}+1)}{2}\ln\left(\frac{\nu_{2}+1}{2}\right)\nonumber \\
 & +\left(\bar{\varepsilon}_{2}+1\right)\ln\left(\bar{\varepsilon}_{2}+1\right)-\left(\bar{\varepsilon}_{2}\right)\ln\left(\bar{\varepsilon}_{2}\right),\label{eq:23}\\
\nonumber 
\end{align}
where we have
\begin{equation}
{\displaystyle \bar{\varepsilon}_{1}=\bar{\varepsilon}_{2}=\frac{1}{4}\left(\left\langle \hat{x}_{a}^{2}\right\rangle +\left\langle \hat{p}_{a}^{2}\right\rangle -2\right)},\label{eq:24}
\end{equation}
\begin{equation}
\nu_{1,2}=\sqrt{\frac{\Delta\pm\sqrt{\Delta^{2}-4|\mathbf{\sigma}|}}{2}},\label{eq:25}
\end{equation}
and
\begin{equation}
\Delta=2\left(\left\langle \hat{x}_{a}^{2}\right\rangle \left\langle \hat{p}_{a}^{2}\right\rangle +\left\langle \hat{x}_{a}\hat{x}_{b}\right\rangle \left\langle \hat{p}_{a}\hat{p}_{b}\right\rangle \right).\label{eq:26}
\end{equation}
Note that the coherence Eq.(\ref{eq:23}) is completely characterized
in terms of the covariance matrix according to section II.

The Wigner function for two particles with state $\hat{\rho}$ can be defined in terms of the characteristic function $\chi(\epsilon)$ as follows,
\begin{equation}
W(R) = \frac{1}{(2\pi)^2} \int_{}^{} d\epsilon \exp{\left[iR^{T}\mathbf{\Lambda} \epsilon \right]}\chi(\epsilon)
\end{equation}
with,
\begin{equation}
R= {\left(\begin{array}{cccc}
q_{a}     & p_{a}     & q_{b}     & p_{b}
      \end{array}\right)}^T
\label{matrix-ISOD}
\end{equation}
\begin{equation}
\hat{\xi}= {\left(\begin{array}{cccc}
\hat{q}_{a}   & \hat{p}_{a}    & \hat{q}_{b}    & \hat{p}_{b}  
      \end{array}\right)}^T
\label{matrix-ISOD}
\end{equation}
\begin{equation}
\epsilon = {\left(\begin{array}{cccc}
u_{1}   & v_{1}  & u_{2}  & v_{2}
      \end{array}\right)}^T
\label{matrix-ISOD}
\end{equation}
\begin{equation}
\mathbf{\Lambda}  = \left(\begin{array}{cccc}
0     & 1     & 0     & 0 \\
-1    & 0     & 0     & 0 \\
0     & 0     & 0     & 1 \\
0     & 0     & -1    & 0 \\
      \end{array}\right)
\label{matrix-ISOD}
\end{equation}
\begin{equation}
\chi(\epsilon) = tr\left[\hat{\rho}\exp{\left(i\epsilon^{T}\mathbf{\Lambda} \hat{\xi}\right)} \right].    
\end{equation}
After some algebraic manipulation, we obtain the Wigner function of the stationary state Eq.(\ref{eq:10-1}),
\begin{align}
\mathcal{W}(x_{a}, p_{a}, x_{b}, p_{b}) &= \frac{\tanh\left(\frac{\Lambda_{+}}{2T} \right) \cdot \tanh\left(\frac{\Lambda_{-}}{2T} \right)}{(\pi)^{2}} \nonumber \\
&\cdot \exp{\left[-\frac{(x_{a} + x_{b})^2}{4(\eta_{1})^2}\tanh\left(\frac{\Lambda_{+}}{2T} \right)\right]} \nonumber \\
&\cdot \exp{\left[-\frac{(p_{a} + p_{b})^2}{4(\eta_{2})^2}\tanh\left(\frac{\Lambda_{+}}{2T} \right) \right]} \nonumber \\ 
&\cdot \exp{\left[-\frac{(x_{a} - x_{b})^2}{4(\eta_{3})^2}\tanh\left(\frac{\Lambda_{-}}{2T} \right)\right]} \nonumber \\
&\cdot \exp{\left[-\frac{(p_{a} - p_{b})^2}{4(\eta_{4})^2}\tanh\left(\frac{\Lambda_{-}}{2T} \right) \right]}
\end{align}
with,
\begin{gather}
\left\{
\begin{array}{lll}
(\eta_{1})^{2} = \frac{\displaystyle \sqrt{\omega + \lambda - \mu}}{\displaystyle 2\omega \sqrt{\omega + \lambda + \mu}} \\
\\
(\eta_{2})^{2} = \frac{\displaystyle \omega \sqrt{\omega + \lambda + \mu}}{ \displaystyle 2 \sqrt{\omega + \lambda - \mu}} 
\end{array}
\right.
\textrm{;}
\left\{
\begin{array}{lll}
(\eta_{3})^{2} = \frac{\displaystyle \sqrt{\omega - \lambda + \mu}}{\displaystyle 2\omega \sqrt{\omega - \lambda - \mu}} \\
\\
(\eta_{4})^{2} = \frac{\displaystyle \omega \sqrt{\omega - \lambda - \mu}}{ \displaystyle 2 \sqrt{\omega - \lambda + \mu}} 
\end{array}
\right. .
\end{gather}

\section{Appendix C: Coherence in the limit of infinite temperature}

In the limit of infinite temperature $T\rightarrow\infty$ the coherence is a function only of the inter-modes coupling parameters and is given by

\begin{equation}
\mathcal{C}(\rho,T\rightarrow\infty)=2\ln\left[\Delta_{1}+\Delta_{2}\right]-\frac{1}{2}\ln\left[\Delta_{3}^{2}-\Delta_{4}^{2}\right],
\end{equation}

with
\begin{equation}
\Delta_{1}=\frac{1}{2(1+\delta_{+})}+\frac{1}{2(1+\delta_{-})},
\end{equation}
\begin{equation}
\Delta_{2}=\frac{\sqrt{\delta_{+}-1}}{4\sqrt{\delta_{-}-1}\sqrt{(\lambda-1)^{2}-\mu^{2}}}+\frac{1}{4\vert1-\delta_{+}\vert},
\end{equation}
\begin{equation}
\Delta_{3}=\frac{4(1-\lambda^{2}-\mu^{2})}{\lambda^{4}+(\mu^{2}-1)^{2}-2\lambda^{2}(\mu^{2}+1)},
\end{equation}
\begin{equation}
\Delta_{4}=-\frac{8\lambda}{\vert\lambda^{4}+(\mu^{2}-1)^{2}-2\lambda^{2}(\mu^{2}+1)\vert},
\end{equation}
\begin{equation}
\delta_{+}=\lambda+\mu \quad \textrm{;} \quad \delta_{-}=\lambda-\mu.
\end{equation}

\end{document}